\documentclass[a4paper,12pt]{article}
\ifx\TeXtures\undefined
\usepackage[psamsfonts]{amsfonts}%
\usepackage{amsmath}%
\else
\usepackage{amsmath}%
\usepackage{amsfonts}%
\usepackage{times}
\fi
\usepackage{amsthm} 
\usepackage{epsfig}         
\usepackage[utf8]{inputenc}
\usepackage[T1]{fontenc}
\usepackage[english]{babel}
\usepackage{ae,icomma} 
\newcommand{\DATUM}{27-04-2021}              
\newcommand{\change}
{{\marginpar{\#}}}        
\newcommand{\comma}{\: ,}              
\newcommand{\period}{\: .}             
\newcommand{\cA}{{\cal A}}
\newcommand{\cB}{{\cal B}}

\newcommand{\cD}{{\cal D}}

\newcommand{\cJ}{{\cal J}}

\newcommand{\cL}{{\cal L}}         

\newcommand{\cW}{{\cal W}}

\newcommand{\field}[1]{\mathbb{#1}}
\newcommand{\R}{\field{R}}            
\newcommand{\Z}{\field{Z}}            
\newcommand{\N}{\field{N}}            
\newcommand{\C}{\field{C}}            

\newcommand{\SSS}{\field{S}}
\newcommand{\rL}{{\rm L}}                 
\newcommand{\rI}{{\rm I}}                 

\newcommand{\rW}{{\rm W}}

\newcommand{\support}{{\rm supp}\, }

\newcommand{\cirS}{\mathop{\bigcirc\kern -.73em {\scriptstyle{\rm S}}}}

\newcommand{\cqfd}{\phantom{blablabla}\hfill\qed\newline} 
\newtheorem{theorem}{Theorem}[section]                
\newtheorem{lemma}[theorem]{Lemma}             
\newtheorem{remark}[theorem]{Remark}           

\theoremstyle{plain}

\newcommand{\donne}{\mapsto}
\newcommand{\dans}{\longrightarrow}

\newcommand{\impl}{\Longrightarrow}

\newcommand{\Pf}{\vspace*{-2mm}{\bf Proof:}\, }

\renewcommand {\l}{\left}
\newcommand {\ri}{\right}

\newcommand{\beq}{\begin{equation}}
\newcommand{\Leq}[1]{\label{#1}\end{equation}}
\newcommand{\eeq}{\end{equation}}
\newcommand{\beqno}{\begin{eqnarray*}}
\newcommand{\eeqno}{\end{eqnarray*}}
\newcommand{\bem}{\l(\! \begin{array}}
\newcommand{\eem}{\end{array}\!\ri)}
\newcommand{\bsm}{\left(\begin{smallmatrix}} 
\newcommand{\esm}{\end{smallmatrix}\right)}  

\begin{document}

\setcounter{section}{0} 

\title{A new proof of the analyticity \\
of the electronic density of molecules.}
\author{
{\bf Thierry Jecko}\\
AGM, UMR 8088 du CNRS, Université de Cergy-Pontoise,\\
Département de mathématiques, site de Saint Martin,\\
2 avenue Adolphe Chauvin,\\
F-95000 Cergy-Pontoise, France. \\
e-mail: thierry.jecko@u-cergy.fr\\
web: http://www.u-cergy.fr/tjecko/\\}
\date{\DATUM}
\maketitle

\begin{abstract}
We give a new, short proof of the regularity away from the nuclei of the electronic 
density of a molecule obtained in \cite{fhhs1,fhhs2}. The new argument is based on the 
regularity properties of the Coulomb interactions underlined in \cite{hu,kmsw}. 
Well-known pseudodifferential techniques for elliptic operators are also used. \\
The paper is published in {\rm Letters in Mathematical Physics} 93, number 1, pp. 73-83, 2010. 
The original publication is available at " www.springerlink.com ". 
\vspace{2mm}

\end{abstract}

\newpage





\section{Introduction.}
\label{intro}
\setcounter{equation}{0}

For the quantum description of molecules, it is very useful to study the so-called electronic density and, in particular, its regularity properties. This has be done for molecules with fixed nuclei: see \cite{fhhs1,fhhs2,fhhs3} for details and references. The smoothness and the analyticity of the density away from the nuclei are proved in \cite{fhhs1} and \cite{fhhs2} respectively. In this paper, we propose an alternative proof. 

Let us recall the framework and the precise results of \cite{fhhs1,fhhs2}. We consider a molecule with $N$ moving electrons ($N\geq 1$) and $L$ fixed nuclei. While the distinct vectors $R_1, \cdots , R_L\in\R^3$ denote the positions of the nuclei, the positions of the electrons are given by $x_1, \cdots , x_N\in\R^3$. The charges of the nuclei are given by the positive $Z_1, \cdots , Z_L$ and the electronic charge is $-1$. In this picture, the Hamiltonian of the system is 
\begin{eqnarray}\label{eq:hamiltonien}
H&:=&\sum _{j=1}^N\Bigl(-\Delta _{x_j}\, -\, \sum _{k=1}^LZ_k|x_j-R_k|^{-1}\Bigr)\, +\, \sum _{1\leq j<j'\leq N}|x_j-x_{j'}|^{-1}\, +\, E_0\comma\\
\mbox{where}\ E_0&=&\sum _{1\leq k<k'\leq L}Z_kZ_{k'}|R_k-R_{k'}|^{-1}\nonumber
\end{eqnarray}
and $-\Delta _{x_j}$ stands for the Laplacian in the variable $x_j$. Setting 
$\Delta :=\sum _{j=1}^N\Delta _{x_j}$, we define the potential $V$ of the system as the multiplication operator satifying $H=-\Delta +V$. Thanks to Hardy's inequality 
\begin{equation}\label{eq:hardy}
\exists c>0\, ;\ \forall f\in \rW^{1,2}(\R^3)\comma \ \int _{\R^3}|t|^{-2}\, |f(t)|^2\, dt\ \leq \ c\int _{\R^3}|\nabla f(t)|^2\, dt\comma 
\end{equation}
one can show that $V$ is $\Delta$-bounded with relative bound $0$ and that $H$ is self-adjoint on the domain of the Laplacian $\Delta $, namely $\rW ^{2,2}(\R^{3N})$ (see Kato's theorem in \cite{rs2}, p. 166-167). If $N< L-1+2\sum _{k=1}^LZ_k$, there exists $E\leq E_0$ 
and $\psi\in \rW ^{2,2}(\R^{3N})\setminus\{0\}$ such that $H\psi =E\psi$ (cf.\ \cite{cfks,fh,rs4}). The electronic density associated to $\psi$ is 
\[\rho (x)\ :=\ \sum _{j=1}^N\int _{\R^{3(N-1)}}\bigl|\psi (x_1,\cdots , x_{j-1}, x, x_j,\cdots , x_N)\bigr|^2\, dx_1\cdots dx_{j-1}dx_j\cdots dx_N\comma \]
an $\rL ^1(\R^3)$-function. For $N=1$, we take $\rho =|\psi |^2$. The regularity result is the following
\begin{theorem}\label{th-anal}{\rm \cite{fhhs1,fhhs2}}.
The density $\rho$ is real analytic on $\R^3\setminus\{R_1,\cdots , R_L\}$.
\end{theorem}
\begin{remark}\label{rk-2preuves}
In \cite{fhhs1}, it is proved that $\rho$ is smooth on $\R^3\setminus\{R_1,\cdots , R_L\}$. This result is then used in \cite{fhhs2} to derive the analyticity. 
\end{remark}
Now let us sketch the new proof of Theorem~\ref{th-anal}, the complete proof and the notation used are given in Section~\ref{details}. We consider the almost everywhere defined $\rL^2$-function 
\begin{equation}\label{tilde-psi}
\tilde\psi \ :\ \R^3\ni x\, \donne \, \psi (x, \cdot ,\cdots , \cdot )\in 
\rW ^{2,2}(\R^{3(N-1)})
\end{equation}
and denote by $\|\cdot \|$ the $\rL^2(\R^{3(N-1)})$-norm. 
By permutation of the variables, it suffices to show that the map $\R^3\ni x\, \donne \, \|\tilde\psi (x)\|^2$ belongs to 
$C^\omega (\R^3\setminus\{R_1,\cdots , R_L\};\R)$, the space of real analytic functions on $\R^3\setminus\{R_1,\cdots , R_L\}$. We define the potentials $V_0$, $V_1$ by 
\begin{equation}\label{eq:def-V_0-V_1}
V\ =\ V_0\, +\, V_1\hspace{.4cm}\mbox{with}\hspace{.4cm}
V_0(x)\ =\ E_0\, -\, \sum _{k=1}^LZ_k|x-R_k|^{-1}\ \in\ C^\omega (\R^3\setminus\{R_1,\cdots , R_L\};\R)\period
\end{equation}
Denoting by $\cB _k :=\cL (\rW ^{k,2}(\R^{3(N-1)});\rL ^2(\R^{3(N-1)}))$ for $k\in\N$, 
\begin{equation}\label{eq:distri}
-\Delta _x\tilde\psi \, +\, Q(x)\tilde\psi  =0\comma \ \mbox{in}\ \cD'(\R^3 ;\rW ^{2,2}(\R^{3(N-1)}))\comma 
\end{equation}
where the $x$-dependent operator $Q(x)\in \cB _2$ is given by $Q(x)=-\Delta _{x'}+V_0-E+V_1$ with $\Delta _{x'}=\sum _{j=2}^N\Delta _{x_j}$.
Considering \eqref{eq:distri} in a small enough, bounded neighbourhood $\Omega$ of some $x_0\in\R^3\setminus\{R_1,\cdots , R_L\}$, we pick from \cite{hu,kmsw} a $x$-dependent 
unitary operator $U_{x}$ on $\rL ^2(\R^{3(N-1)})$ such that 
\begin{equation}\label{eq:potentiel-analytique}
W:\ \Omega\ni x\donne U_{x}V_1U_{x}^{-1}\in \cB _1\, \subset\, \cB _2
\end{equation}
is analytic. It turns out that $P_0=U_{x}(-\Delta _x-\Delta _{x'})U_{x}^{-1}$ is an elliptic differential operator in the variable $(x, y)$ but can be considered as a differential operator in $x$ with analytic, differential coefficients in $\cB _2$. Applying $U_{x}$ to \eqref{eq:distri} and setting $\varphi (x)=U_{x}\tilde\psi (x)$, we obtain 
\begin{equation}\label{eq:distri-u-x_0}
(P_0\, +\, W\, +\, V_0\, -\, E)\, \varphi \ =\ 0\period
\end{equation}
Since $U_{x}$ is unitary on $\rL ^2(\R^{3(N-1)})$, $\|\tilde\psi (x)\|=\|\varphi (x)\|$. It suffices to prove that $\varphi \in C^\omega (\Omega ;\rL^2(\R^{3(N-1)}))$.
Using \eqref{eq:distri-u-x_0} and a parametrix of the elliptic operator $P_0$, we show that, for all $k$, $\varphi\in \rW^{k,2}(\Omega ;\rW^{1,2}(\R^{3(N-1)}))$ by induction and, using the same tools again, that $\varphi\in \rW^{k,2}(\Omega ;\rW^{2,2}(\R^{3(N-1)}))$, for all $k$. Thus $\varphi\in C^\infty(\Omega ;\rW^{2,2}(\R^{3(N-1)}))$. Viewing $P_0+W+V_0$ as a differential operator in $x$, we can adapt the arguments in \cite{h1} p. 178-180 to get $\varphi\in C^\omega (\Omega ;\rW^{2,2}(\R^{3(N-1)}))$, yielding $\varphi\in C^\omega (\Omega ;\rL^2(\R^{3(N-1)}))$.

The main idea in the construction of the unitary operator $U_{x}$ is to change, locally in $x$, the variables $x_2,\cdots , x_N$ in a $x$-dependent way such 
that the $x$-dependent singularities $|x-x_j|^{-1}$ becomes locally $x$-independent (see Section~\ref{details}). In \cite{hu}, where this clever method was introduced, and in 
\cite{kmsw}, the nuclei positions play the role of the $x$ variable and the $x_2,\cdots , x_N$ are the electronic degrees of freedom. In \cite{kmsw}, the accuracy of the Born-Oppenheimer approximation is proved for the computation of the eigenvalues and eigenvectors of the molecule. We point out that this method is the core of a 
semiclassical pseudodifferential calculus adapted to the treatment of Coulomb singularities in molecular systems, namely the twisted $h$-pseudodifferential calculus ($h$ being the semiclassical parameter). This calculus is due to A. Martinez and V. Sordoni in \cite{ms}, where the Born-Oppenheimer approximation for molecular time evolution is validated. 

As one can see in \cite{kmsw,ms}, the above method works in a larger framework. 
So do Theorem~\ref{th-anal} and our proof. For instance, we do not need the positivity of the charges $Z_k$, the fact that $E\leq E_0$, and the precise form of the Coulomb interaction. We do not use the self-adjointness (or the symmetry) of the 
operator $H$. We could replace in \eqref{eq:hamiltonien} each $-\Delta _{x_j}$ by $|i\nabla _{x_j}+A(x)|^2$, where $A$ is a suitable, analytic, magnetic vector potential. We could also add a suitable, analytic exterior potential. 

Let us now compare our proof with the one in \cite{fhhs1,fhhs2}. Here we use known arguments of elliptic regularity (cf. \cite{h1}). This is also the case in 
\cite{fhhs1,fhhs2}. In those papers however, the authors directly show the regularity of $\psi$ in some appropriate directions and use it in the formula for $\rho$ with the help of a smartly chosen partition of unity. Here the $x$-dependent change of variables produces regularity with respect to $x$. As external tools, we only exploit basic facts of pseudodifferential calculus, the rest being elementary. We believe that, in spirit, the two proofs are similar. \\
We note that the clever method borrowed from \cite{hu, kmsw}, which transforms the singular potential $V_1$ in an analytic function with values in $\cB_1$, allows us to 
treat the regularity problem with known technics of elliptic regularity. 

{\bf Acknowledgment:} The author is supported by the french ANR grant ``NONAa'' and by the european GDR ``DYNQUA''. He thanks Vladimir Georgescu, Sylvain Golénia, Hans-Henrik Rugh, and Mathieu Lewin, for stimulating discussions.

\section{Details of the proof.}
\label{details}
\setcounter{equation}{0}

Here we complete the proof of Theorem~\ref{th-anal}, sketched in Section~\ref{intro}. 

{\bf Notation and basic facts.} For a function $f:\R^d\times\R^n\ni (x,y)\donne f(x,y)\in\R^p$, let $d_xf$ be the total derivative of $f$ w.r.t.\ $x$, by $\partial _x^\alpha f$ with $\alpha\in\N^d$ the corresponding partial derivatives. For $\alpha\in\N^d$ and $x\in\R^d$, $D_x^\alpha :=(-i\partial _x)^\alpha :=(-i\partial _{x_1})^{\alpha _1}\cdots(-i\partial _{x_d})^{\alpha _d}$, $D_x=-i\nabla_x$, $x^\alpha :=x_1^{\alpha_1}\cdots x_d^{\alpha_d}$, $|\alpha |:=\alpha_1+\cdots +\alpha _d$, $\alpha !:=(\alpha_1!)\cdots (\alpha_d!)$, $|x|^2=x_1^2+\cdots +x_d^2$, and $\langle x\rangle :=(1+|x|^2)^{1/2}$. If $\cA$ is a Banach space and $O$ an open subset of $\R^d$, we denote by $C_c^\infty(O;\cA)$ (resp. $C_b^\infty(O;\cA)$, resp. $C^\omega(O;\cA)$) the space of functions from $O$ to $\cA$ which are smooth with compact support (resp. smooth with bounded derivatives, resp. analytic).
Let $\cD '(O;\cA)$ denotes the topological dual of $C_c^\infty(O;\cA)$. We use the traditional notation $\rW^{k,2}(O;\cA)$ for the Sobolev spaces of $\rL^2(O;\cA)$-functions with $k$ derivatives in $\rL^2(O;\cA)$ when $k\in\N$ and for the dual of $\rW^{-k,2}(O;\cA)$ when $-k\in\N$. If $\cA'$ is another Banach space, we denote by $\cL (\cA ;\cA')$ the space of the continuous linear maps from $\cA$ to $\cA'$ and set $\cL (\cA)=\cL (\cA ;\cA)$. For $A\in\cL (\cA)$ with finite dimensional $\cA$, $A^{\mathrm T}$ denotes the transpose of $A$ and ${\rm Det}A$ its determinant. By the Sobolev injections, 
\begin{equation}\label{sobolev}
\bigcap _{k\in\N}\rW^{k,2}(O;\cA)\ \subset \ C^\infty (O;\cA)\period
\end{equation}
Let $\|\cdot \|_\cA$ be the norm of $\cA$ and let $\delta\in\{0;1\}$. Recall (cf.\ the appendix) that a function $u\in C^\infty(O;\cA)$ is real analytic if and only if, for any compact $K\subset O$, there exists $\delta\in\{0; 1\}$ and $A_\delta>0$ such that 
\begin{equation}\label{caract-anal}
 \forall \alpha\in\N^d\comma \hspace{.4cm} \sup _{x\in K}\, 
\bigl\|(D_x^\alpha u)(x)\bigr\|_{\cA}\ \leq \ A_\delta^{|\alpha |+1}\cdot (\alpha !)^\delta\cdot (|\alpha |!)^{1-\delta}\period
\end{equation}
For convenience, we set $\cW_k=\rW^{k,2}(\R^{3(N-1)})$, for $k\in\N$. Recall that $\cB _k=\cL (\cW_k; \cW_0)$. 

{\bf Construction of $U_{x}$ (see \cite{hu,kmsw,ms}).} Let $\tau\in C_c^\infty 
(\R^3 ;\R)$ with $\tau (x_0)=1$ and $\tau =0$ near $R_k$, for all $k\in \{1;\cdots ; L\}$. For $x, s\in\R^3$, let $f(x, s)=s+\tau (s)(x-x_0)$. 
\begin{equation}\label{eq:prop-f}
\mbox{Notice that}\hspace{.4cm} f(x, x_0)=x \hspace{.4cm}\mbox{and}\hspace{.4cm}f(x, s)=s \hspace{.4cm}\mbox{if}\hspace{.4cm} s\not\in\support\tau\period
\end{equation}
Since $(d_sf)(x, s)\cdot s'=s'+\langle\nabla \tau (s), s'\rangle (x-x_0)$, we can choose a small enough, relatively compact neighbourhood $\Omega$ of $x_0$ such that 
\begin{equation}\label{eq:diffeo}
\forall x\in\Omega\comma \hspace{.4cm}\sup _{s}\|(d_sf)(x, s)\, -\, \rI _3\|_{\cL (\R^3)}\ \leq \ 1/2\comma
\end{equation}
$\rI _3$ being the identity matrix of $\cL (\R^3)$. Thus, for $x\in\Omega$, $f(x, \cdot )$ is a $C^\infty$-diffeomorphism on $\R^3$ and we denote by $g(x, \cdot )$ its inverse. By \eqref{eq:diffeo} and a Neumann expansion in $\cL (\R^3)$, 
\[\bigl((d_sf)(x, s)\bigr)^{-1}\ =\ \rI _3\, +\, \Bigl(\sum _{n=1}^\infty \bigl(-\langle\nabla \tau (s), (x-x_0)\rangle\bigr)^{n-1}\Bigl)\langle\nabla \tau (s), \cdot\rangle (x-x_0)\comma\]
for $(x, s)\in\Omega\times\R^3$. Notice that the power series converges uniformly w.r.t.\ $s$. This is still true for the series of the derivatives $\partial _s^\beta$, for $\beta\in\N^3$. Since
\begin{equation}
(d_sg)(x, f(x, s))=\bigl((d_sf)(x, s)\bigr)^{-1}\hspace{.2cm}\mbox{and}\hspace{.2cm}
(d_xg)(x, f(x, s))=-\tau (s)(d_sg)(x, f(x, s))\comma
\end{equation}
we see by induction that, for $\alpha , \beta\in\N^3$, 
\begin{equation}\label{eq:diffeo-anal-g}
\bigl(\partial_x^\alpha\partial _s^\beta g\bigr)(x, f(x, s))\ =\ \sum _{\gamma\in\N^3}(x-x_0)^\gamma a_{\alpha\beta\gamma}(s)
\end{equation}
on $\Omega\times\R^3$, with coefficients $a_{\alpha\beta\gamma}\in C^\infty (\R^3;\cL (\R^3))$. For $\alpha =\beta=0$, this follows from $g(x, f(x, s))=s$. 
Notice that, except for $(\alpha , \beta , \gamma )=(0, 0, 0)$ and for $|\beta |=1$ with $(\alpha , \gamma )=(0, 0)$, the coefficients $a_{\alpha\beta\gamma}$ are supported in the compact support of $\tau$. \\
For $x\in \R^3$ and $y=(y_2,\cdots , y_N)\in\R^{3(N-1)}$, let $F(x, y)=(f(x, y_2),\cdots , f(x, y_N))$. For $x\in\Omega$, $F(x, \cdot )$ is a $C^\infty$-diffeomorphism on $\R^{3(N-1)}$ satisfying the following properties: There exists $C_0>0$ such that, for all $\alpha\in\N^{3}$, for all $x\in\Omega$, for all $s, s'\in\R^{3}$, 
\begin{eqnarray}
C_0^{-1}|s-s'|\, \leq \, |f(x, s)-f(x, s')|\, \leq \, C_0|s-s'|\comma\label{eq:prop-F-1}\\
|\partial _x^\alpha f(x, s)-\partial _x^\alpha f(x, s')|\, \leq \, C_0|s-s'|\comma\label{eq:prop-F-2}\\
\mbox{and, for}\, |\alpha |\geq 1\comma \ |\partial _x^\alpha f(x, s)|\, \leq \, C_0\period\label{eq:prop-F-3}
\end{eqnarray}
For $x\in\Omega$, denote by $G(x, \cdot)$ the inverse diffeomorphism of $F(x, \cdot )$. By \eqref{eq:diffeo-anal-g}, the functions $\Omega\times\R^{3(N-1)}\ni (x,y)\donne (\partial_x^\alpha\partial _y^\beta G)(x, F(x, y))$, 
%
%
for $(\alpha , \beta )\in\N^3\times\N^{3(N-1)}$, are also given by a power series in $x$ with smooth coefficients in $y$. Given $x\in\Omega$, let $U_{x}$ be the unitary operator on $\rL^2(\R^{3(N-1)})$ defined by
\begin{equation}\label{eq:def-U}
(U_{x}\theta )(y)=|{\rm Det}(d_yF)(x, y)|^{1/2}\theta (F(x, y))\period
\end{equation}
{\bf Computation of the terms in \eqref{eq:distri-u-x_0} (cf.\ \cite{kmsw,ms}).} 
Consider the functions 
\begin{eqnarray*}
\Omega\ni x\donne J_1(x, \cdot )&\in &C^\infty_b \bigl(\R^{3(N-1)};\cL (\R^{3(N-1)};\R^3)\bigr)\comma\\
\Omega\ni x\donne J_2(x, \cdot )&\in &C^\infty _b(\R^{3(N-1)};\R^3)\comma\\
\Omega\ni x\donne J_3(x, \cdot )&\in &C^\infty _b\bigl(\R^{3(N-1)};\cL (\R^{3(N-1)})\bigr)\comma\\
\Omega\ni x\donne J_4(x, \cdot )&\in &C^\infty _b(\R^{3(N-1)};\R^{3(N-1)})\comma\\
\mbox{defined by }\ J_1(x, y)&=&(d_xG(x, y'))^{\mathrm{T}}\bigl(x,\, y'=F(x, y)\bigr)\comma \\
J_2(x, y)&=&\bigl|{\rm Det}\, d_yF(x, y)\bigr|^{1/2}\, D_x\Bigl(\bigl|{\rm Det}\, d_{y'}G(x, y')\bigr|^{1/2}\Bigr)\Bigr|_{y'=F(x, y)}\comma\\
J_3(x, y)&=&(d_{y'}G(x, y'))^{\mathrm{T}}\bigl(x,\, y'=F(x, y)\bigr)\comma \\
J_4(x, y)&=&\bigl|{\rm Det}\, d_yF(x, y)\bigr|^{1/2}\, D_{y'}\Bigl(\bigl|{\rm Det}\, d_{y'}G(x, y')\bigr|^{1/2}\Bigr)\Bigr|_{y'=F(x, y)}\period 
\end{eqnarray*} 
Thanks to \eqref{eq:diffeo-anal-g}, the $J_k(\cdot , y)$'s can also be written as a power series in $x$ with smooth coefficients depending on $y$. Now 
\begin{eqnarray}\label{eq:nabla-conj}
U_{x}\nabla _xU_{x}^{-1}\ =\ \nabla _x\, +\, J_1\nabla _y\, +\, J_2\comma\hspace{.4cm}U_{x}\nabla _{x'}U_{x}^{-1}\ =\ J_3\nabla _y\, +\, J_4\comma\hspace{.4cm}\mbox{and}\hspace{.4cm}\\
\label{eq:delta-conj}
P_0\ =\ U_{x}\bigl(-\Delta _x\, -\, \Delta _{x'}\bigr)U_{x}^{-1}\ =\ -\Delta _x\, +\, \cJ _1(x; y; D_y)\cdot D_x\, +\, \cJ _2(x; y; D_y)\comma 
\end{eqnarray}
where $\cJ _2(x; y; D_y)$ is a scalar differential operator of order $2$ and $\cJ _1(x; y; D_y)$ is a column vector of $3$ scalar differential operators of order $1$. Actually the coefficients of $\cJ _1(x; y; D_y)$ and of $\cJ _2(x; y; D_y)$ belong to $C_b^\infty(\Omega \times \R^{3(N-1)}; \C)$. 
By \eqref{eq:diffeo-anal-g}, $\cJ _1$ (resp. $\cJ _2$) is given on $\Omega$ by a power series of $x$ with coefficients in $\cB _1$ (resp. $\cB _2$) and therefore is a real analytic function on $\Omega$ with values in $\cB_1$ (resp. $\cB _2$) (cf.\ \cite{h3}). Next, we look at $W$ defined in \eqref{eq:potentiel-analytique}. By \eqref{eq:prop-f} and \eqref{eq:def-U},  $j\neq j'$ in $\{2;\cdots ; N\}$, for $k\in\{1;\cdots ; L\}$, and for $x\in\Omega$,
\begin{eqnarray}
U_{x}\bigl(|x-x_j|^{-1}\bigr)U_{x}^{-1}&=&|f(x; x_0)-f(x; y_j)|^{-1}\comma \label{eq:pot-conj-1}\\
U_{x}\bigl(|x_j-R_k|^{-1}\bigr)U_{x}^{-1}&=&|f(x; y_j)-f(x; R_k)|^{-1}\comma
\label{eq:pot-conj-2}\\
U_{x}\bigl(|x_j-x_{j'}|^{-1}\bigr)U_{x}^{-1}&=&|f(x; y_j)-f(x; y_{j'})|^{-1}
\period\label{eq:pot-conj-3}
\end{eqnarray}
\begin{lemma}\label{l:pot-anal}
The potential $W$, defined in \eqref{eq:potentiel-analytique}, is an real analytic function on $\Omega$ with values in 
$\cB _1=\cL (\cW _1, \cW_0)$. 
\end{lemma}
\Pf Notice that $W$ is a sum of terms of the form \eqref{eq:pot-conj-1}, \eqref{eq:pot-conj-2}, and \eqref{eq:pot-conj-3}. We show the regularity of \eqref{eq:pot-conj-1}. Similar arguments apply for the other terms. We first recall the arguments in \cite{kmsw}, which proves the $C^\infty$ regularity.\\
Using the fact that $d_x(f(x, x_0)-f(x, y_j))$ does not depend on $x$, 
\[D_x^\alpha \bigl(|f(x, x_0)-f(x, y_j)|^{-1}\bigr)\ =\ (\tau (x_0)-\tau (y_j))^{|\alpha |}\, \bigl(D^\alpha|\cdot |^{-1}\bigr)(f(x, x_0)-f(x, y_j)) \]
for $x_0\neq y_j$. It is straightforward to check that 
\begin{equation}\label{eq:facile-derivees-successives}
\forall \alpha\in\N^3\comma\ \exists C>0\comma\ \forall y\in\R^3\setminus\{0\}\comma\hspace{.4cm} \bigl|D^\alpha |\cdot |^{-1}\bigr|(y)\ \leq \ C(\alpha !)\, |y|^{-|\alpha |-1}\period 
\end{equation}
By \eqref{eq:prop-F-1}, \eqref{eq:prop-F-2} with $|\alpha |=1$, and \eqref{eq:facile-derivees-successives}, we see that, for all $\alpha\in\N^3$ and for $x_0\neq y_j$, 
\begin{eqnarray*}
 \bigl|D_x^\alpha \bigl(|f(x, x_0)-f(x, y_j)|^{-1}\bigr)\bigr|&\leq& 
C_0^{2|\alpha |}|f(x, x_0)-f(x, y_j)|^{|\alpha |}\, \bigl|D^\alpha|\cdot |^{-1}\bigr|(f(x, x_0)-f(x, y_j)) \\
&\leq& C_0^{2|\alpha |}C(\alpha !)\cdot |f(x, x_0)-f(x, y_j)|^{-1}\\
&\leq&C_0^{2|\alpha |}C(\alpha !)C_0\cdot |x_0-y_j|^{-1}\comma
\end{eqnarray*}
where we used again \eqref{eq:prop-F-1} in the last inequality. Thus, by \eqref{eq:hardy}, 
\begin{equation}\label{eq:derivees-successives-bornees}
\bigl\|D_x^\alpha \bigl(|f(x, x_0)-f(x, y_j)|^{-1}\bigr)\bigr\|_{\cB _1}\ \leq \  (cCC_0)C_0^{2|\alpha |}(\alpha !)\comma 
\end{equation}
uniformly w.r.t.\ $\alpha\in\N^3$ and $x\in\Omega$. Therefore $W$ is a distribution on $\Omega$ the derivatives of which belong to $\rL^\infty(\Omega)$, thus to $\rL^2(\Omega)$. By \eqref{sobolev}, $W$ is smooth. \\
Using the following improvement of \eqref{eq:facile-derivees-successives}, proved in appendix  below, 
\begin{equation}\label{eq:derivees-successives}
\exists K>0\, ;\ \forall \alpha\in\N^3\comma\ \forall y\in\R^3\setminus\{0\}\comma\hspace{.4cm} \bigl|D^\alpha |\cdot |^{-1}\bigr|(y)\ \leq \ K^{|\alpha |+1}(\alpha !)\, |y|^{-|\alpha |-1}\comma
\end{equation}
the l.h.s.\ of \eqref{eq:derivees-successives-bornees} is, for $\alpha\in\N^3$ and $x\in\Omega$, bounded above by $cC_0C_0^{2|\alpha |}K^{|\alpha |+1}(\alpha !)\leq K_1^{|\alpha |+1}(\alpha !)$, for some $K_1>0$. This yields the result by \eqref{caract-anal} with $\delta =1$. \cqfd

{\bf Smoothness.} We would like to see \eqref{eq:distri-u-x_0} as an ``elliptic'' differential equation w.r.t. $x$ with coefficients in $\cB _2$ and follow usual arguments of elliptic regularity to prove the smoothness of $\varphi$. It turns out that the ellipticity w.r.t $x$ is not well suited to this purpose. Instead, we shall use the ellipticity in all variables of $P_0$. \\
Using \eqref{eq:nabla-conj}, we see that the principal symbol of $P_0$ is given on $\Omega\times \R^{3(N-1)}\times \R^{3N}$ by 
\begin{eqnarray}
p_2(x, y; \xi , \eta)&=& |\xi |^2\, +\, 2\, \langle J_1(x, y)\eta \, ,\, \xi\rangle \, +\, |J_1(x, y)\eta|^2\, +\, |J_3(x, y)\eta|^2 \nonumber\\
&=&|\xi \, +\, J_1(x, y)\eta|^2\, +\, |J_3(x, y)\eta|^2\period\label{eq:princ-symb}
\end{eqnarray}
We observe that there exist $M_1, M_3>0$ such that, for all $(x, y)\in \Omega\times\R^{3(N-1)}$, 
\[\bigl\|J_1(x, y)\bigr\|_{\cL (\R^{3(N-1)};\R^3)}\ \leq \ M_1\hspace{.4cm}\mbox{and}\hspace{.4cm}\bigl\|J_3(x, y)^{-1}\bigr\|_{\cL (\R^{3(N-1)})}\ \leq \ M_3\period\]
We notice that $|J_3(x, y)\eta|\geq M_3^{-1}|\eta|$. Let $S=\sqrt{1+4M_1^2}$. Consider first the case where $S|\eta|\leq (|\xi|^2+|\eta |^2)^{1/2}$. We have $2M_1|\eta|\leq\ |\xi|$. Thus 
\[\bigl|\xi\, +\, J_1(x ; y)\eta\bigr|^2\ \geq \ \frac{|\xi|^2}{4}\]
and, using \eqref{eq:princ-symb}, we obtain the lower bound $p_2(x, y; \xi , \eta)\geq \min (1/4; M_3^{-2})(|\xi|^2+|\eta |^2)$. 
If, now, $S|\eta|\geq (|\xi|^2+|\eta |^2)^{1/2}$, it follows from \eqref{eq:princ-symb} that $p_2(x, y; \xi , \eta)\geq (M_3S)^{-2}(|\xi|^2+|\eta |^2)$. This yields the ellipticity of $P_0$. \\
Let $\chi\in C_c^\infty(\R^3)$ supported in $\Omega$ such that $\chi =1$ near $x_0$. We consider the following elliptic extension of $P_0$: 
\begin{equation}\label{tildeP_0}
 \tilde{P}_0\ =\ -\Delta _x\, +\, \chi (x)\cJ _1(x; y; D_y)\cdot D_x\, +\, \chi ^2(x)
\cJ _2(x; y; D_y)\, +\, (1-\chi^2)(x)(-\Delta _y)\period
\end{equation}
For $m\in\Z$, the class $S^m$ in \cite{h2} (p. 65-75) is the set of smooth functions $a$ on $\R^{6N}$ such that, for all $(\alpha , \beta )\in (\N^{3N})^2$, there exists $C_{\alpha , \beta}>0$ such that, for all $(x, y; \xi , \eta)$, 
\begin{equation}\label{symbole}
 (1+|\xi |^2+|\eta |^2)^{|\beta |/2}|\partial _{x, y}^\alpha\partial 
_{\xi , \eta}^{\beta}a(x, y; \xi , \eta)|\ \leq \ C_{\alpha , \beta}(1+|\xi |^2+|\eta |^2)^{m/2}\period
\end{equation}
Notice that $\tilde{P}_0=\tilde{p}_2(x, y; D_x , D_y)+\tilde{p}(x, y; D_x , D_y)$ with $\tilde{p}\in S^1$ and principal symbol $\tilde{p}_2\in S^2$. Using the ellipticity of $P_0$, one can verify that there exists $C>0$ such that, for $(x, y, \xi , \eta)\in (\R^{3N})^2$ with $|\xi |^2+|\eta |^2\geq 1$, $\tilde{p}_2\geq C(|\xi |^2+|\eta |^2)$. Let $\theta\in C_c^\infty (\R^{3N})$ such that $\theta (\xi , \eta)=1$ if $|\xi |^2+|\eta |^2\leq 1$. Then we see that $q(x, y; \xi , \eta):=(1-\theta (\xi , \eta))(\tilde{p}_2(x, y; \xi , \eta))^{-1}$ belongs to $S^{-2}$. By the composition properties of this pseudodifferential calculus (see \cite{h2} p. 65-75), for some symbols $r_0, r_1, r\in S^{-1}$, 
\begin{eqnarray*}
q(x, y; D_x , D_y)\tilde{P}_0&=&q(x, y; D_x , D_y)\tilde{p}_2(x, y; D_x , D_y)\, +\, r_0(x, y; D_x , D_y)\\
&=&(q\tilde{p}_2)(x, y; D_x , D_y)\, +\, r_1(x, y; D_x , D_y)\ =\ I\, +\, r(x, y; D_x , D_y)\period
\end{eqnarray*}
Setting $Q=q(x, y; D_x , D_y)$ and $R=r(x, y; D_x , D_y)$, we obtain, for all $k\in\N$, 
\begin{eqnarray}
Q\tilde{P}_0&=&I\, +\, R\comma\label{parametrice}\\
Q\, \in\, \cL\bigl(\rW^{k, 2}(\R^{3N}); \rW^{k+2, 2}(\R^{3N})\bigr)\comma &\mbox{and}&R\, \in\, \cL \bigl(\rW^{k, 2}(\R^{3N}); \rW^{k+1, 2}(\R^{3N})\bigr)\comma\label{regularisation}
\end{eqnarray}
by the boundedness properties of this calculus on Sobolev spaces (see \cite{h2} p. 65-75). 
Let $\chi _0\in C_c^\infty(\R^3)$ with $\chi _0=1$ near $x_0$ and $\chi\chi_0=\chi_0$. Applying \eqref{parametrice} to $\chi _0\varphi$, we get $\chi _0\varphi=-R\chi _0\varphi+Q\tilde{P}_0\chi _0\varphi$. Since $\tilde{P}_0\chi _0\varphi = [\tilde{P}_0, \chi _0]\varphi +\chi _0P_0\varphi =[\tilde{P}_0, \chi _0]\chi\varphi+(E-V_0-W)\chi _0\varphi $, 
\begin{equation}\label{elliptique-chi}
\chi _0\varphi \ =\ -R\chi _0\varphi\, +\, Q(E-V_0)\chi _0\varphi \, -\, QW\chi _0\varphi\, +\, Q[\tilde{P}_0, \chi _0]\chi\varphi  \period
\end{equation}
Recall that $\psi\in\rW ^{2, 2}(\R^{3N})$. By \eqref{eq:nabla-conj}, $\chi\varphi =\chi U_x\psi\in\rW ^{2, 2}(\R^{3N})$. In particular, $\chi\varphi , \chi_0\varphi\in\rW ^{1, 2}(\R^{3}; \cW_1)$. By \eqref{regularisation}, $R\chi _0\varphi\in\rW ^{2, 2}(\R^{3}; \cW_1)$ and $Q(E-V_0)\chi _0\varphi\in\rW ^{3, 2}(\R^{3}; \cW_1)$ thanks to \eqref{eq:def-V_0-V_1}. By 
Lemma~\ref{l:pot-anal}, $W\chi _0\varphi\in\rW ^{1, 2}(\R^{3}; \cW_0)$ but $QW\chi _0\varphi\in\rW ^{2, 2}(\R^{3}; \cW_1)$ by \eqref{regularisation}. By \eqref{tildeP_0}, 
$[\tilde{P}_0, \chi _0]\chi\varphi\in\rW ^{0, 2}(\R^{3}; \cW_1)+\rW ^{1, 2}(\R^{3}; \cW_0)$ thus 
$Q[\tilde{P}_0, \chi _0]\chi\varphi\in\rW ^{2, 2}(\R^{3}; \cW_1)$. Now 
\eqref{elliptique-chi} implies that $\chi_0\varphi\in\rW ^{2, 2}(\R^{3}; \cW_1)$. 
Using this new information and a cut-off $\chi _1\in C_c^\infty(\R^3)$ such that $\chi _1=1$ near $x_0$ and $\chi_0\chi_1=\chi_1$, we get in the same way, $\chi$ (resp. $\chi _0$) being replaced by $\chi _0$ (resp. $\chi _1$), that $\chi_1\varphi\in\rW ^{3, 2}(\R^{3}; \cW_1)$. So, by induction, $\varphi\in\rW ^{k, 2}(\Omega '; \cW_1)$, for all $k\in\N$, on some neighbourhood $\Omega '$ of $x_0$. By \eqref{sobolev}, $\varphi\in C^\infty (\Omega ';\cW_1)$. 

{\bf Remarks:} We have recovered the result in \cite{fhhs1}. To get it, we needed neither the refined bounds \eqref{eq:derivees-successives} nor the power series mentioned above but just used the smoothness of $f$ w.r.t.\ $x$. \\
Starting from $\chi\varphi\in\rW^{k, 2}(\R^{3}; \cW _1)$, for some $k\in\N$, $W\chi _0\varphi\in\rW ^{k,2}(\R^{3}; \cW_0)$ by Lemma~\ref{l:pot-anal}. Now we use \eqref{regularisation} to see that $R\chi _0\varphi, QW\chi _0\varphi, Q[\tilde{P}_0, \chi _0]\chi\varphi\in\rW ^{k, 2}(\R^{3}; \cW_2)$, yielding $\chi _0\varphi\in\rW ^{k,2}(\R^{3}; \cW_2)$ by \eqref{elliptique-chi}. Therefore $\varphi\in C^\infty (\R^3\setminus\{R_1, \cdots , R_L\};\cW_2)$. \\
We could have used a local pseudodifferential calculus (cf. \cite{h2} p. 83-87) and wave front sets (cf. \cite{h2} p. 88-91) to get a more elegant but more involved proof. We proved \eqref{parametrice} which is a very weak version of the ellipticity result in \cite{h2}, p. 72-73. For the non specialists' sake, we prefered to use elementary tools, admiting only the results on composition and on boundedness on Sobolev spaces of the basic pseudodifferential calculus given in \cite{h2}, p. 65-76. 

{\bf Analyticity.} By the second remark above, we know that $\varphi \in C^\infty (\Omega;\cW _2)$. To show that $\varphi \in C^\omega (\Omega;\cW _2)$, we adapt the proof of Theorem 7.5.1 in \cite{h1} for equation \eqref{eq:distri-u-x_0}. So we view the latter as $P\varphi =0$ where $P=\sum _{|\alpha |\leq 2}a_\alpha D_x^\alpha$ with analytic differential $\cB _{2-|\alpha|}$-valued coefficients $a_\alpha$ (cf.\ Lemma~\ref{l:pot-anal}, \eqref{eq:def-V_0-V_1}, and \eqref{eq:delta-conj}). Because of the low regularity in $y$, we essentially follow the proof of Lemma 3.1 in \cite{fhhs2}. \\
Take $\chi $ and $\Omega '$ as in the proof of the smoothness of $\rho$ and with  
$\chi=1$ on $\Omega '$. We shall prove that $\varphi \in C^\omega (\Omega ';\cW _2)$. To this end, we strengthen a little bit \eqref{parametrice}. Let $Q_1=(I-R)Q$. Then $Q_1=q_1(x, y; D_x , D_y)$ with $q_1\in S^{-2}$ and, for some $\tilde{r}\in S^{-2}$, 
\begin{eqnarray}\label{parametrice-2}
Q_1\tilde{P}_0\ =\  (I-r(x, y; D_x , D_y))(I+r(x, y; D_x , D_y))\ =\ I - \tilde{r}(x, y; D_x , D_y)\comma\\
Q_1\comma \, R_1\, :=\, \tilde{r}(x, y; D_x , D_y)\, \in\,  \cL\bigl(\rW^{k, 2}(\R^{3N}); \rW^{k+2, 2}(\R^{3N})\bigr)\period\label{regularisation-2}
\end{eqnarray}
We claim that there exists $C>0$ such that, for all $v\in C_c^\infty (\Omega ';\cW _2)$, $r\in\{0;1;2\}$, $\alpha\in\N^3$,
\begin{equation}\label{eq:borne-elliptique}
|\alpha |+r\leq 2\ \impl \ \|D_x^\alpha v\|_{\rL ^2(\Omega ';\cW _r)}\ \leq \ C\|Pv\|_{\rL ^2(\Omega ';\cW _0)}\, +\, C\|v\|_{\rL ^2(\Omega ';\cW _0)}\period
\end{equation}
By \eqref{parametrice-2} and \eqref{regularisation-2}, we see that \eqref{eq:borne-elliptique} holds true if $P$ is replaced by $\tilde{P}_0$. Since $\tilde{P}_0v=P_0v$ if $v\in C_c^\infty (\Omega ';\cW _2)$, \eqref{eq:borne-elliptique} holds true if $P$ is replaced by $P_0$.
Recall that $P=P_0+W+V_0-E$. Since $V$ and $V_0$ are $(\Delta _x+\Delta _{x'})$-bounded with relative bound $0$, $W$ is $P_0$-bounded with relative bound $0$, by the properties of $U_{x}$. This means in particular that there exists $C'>0$ such that, for all $v\in C_c^\infty (\Omega ';\cW _2)$, 
\[\|(W+V_0-E) v\|_{\rL ^2(\Omega ';\cW _0)}\ \leq \ (1/2)\|P_0v\|_{\rL ^2(\Omega ';\cW _0)}\, +\,  C'\|v\|_{\rL ^2(\Omega ';\cW _0)}\period\]
For such $v$, $\|P_0v\|_{\rL ^2(\Omega ';\cW _0)}\leq \|Pv\|_{\rL ^2(\Omega ';\cW _0)}+(1/2)\|P_0v\|_{\rL ^2(\Omega ';\cW _0)}+C'\|v\|_{\rL ^2(\Omega ';\cW _0)}$. Thus \eqref{eq:borne-elliptique} follows from the same estimate with $P$ replaced by $P_0$.\\
For $\epsilon>0$, let $\Omega _\epsilon ':=\{x\in\Omega';\, d(x;\R^3\setminus\Omega ')>\epsilon\}$ and, for $r\in\N$, denote the $\rL^2(\Omega _\epsilon ';\cW _r)$-norm of $v$ by $N_{\epsilon , r}(v)$. As in \cite{h1} (Lemma 7.5.1), we use an appropriate cut-off function, Leibniz' formula, and \eqref{eq:borne-elliptique}, to find $C_e>0$ such that, for all $v\in C^\infty (\Omega ';\cW _2)$, for all $\epsilon , \epsilon _1\geq 0$, for all $r\in\{0;1;2\}$ and all $\alpha\in\N^3$ such that $r+|\alpha |\leq 2$, 
\begin{equation}
\epsilon ^{r+|\alpha |}N_{\epsilon +\epsilon_1, r}(D_x^\alpha v)\ \leq \ C_e\epsilon ^2N_{\epsilon_1, 0}(Pv)\, +\, C_e\sum_{r+|\alpha '|<2}\, 
\epsilon ^{r+|\alpha '|}N_{\epsilon_1, r}(D_x^{\alpha '}v)\period\label{eq:borne-elliptique-N}
\end{equation}
We used the fact that \eqref{eq:borne-elliptique-N} holds true for $\epsilon >D'$, the diameter of $\Omega'$, since the l.h.s.\ is  zero. By \eqref{caract-anal} with $\delta =0$, there exists $C_p>0$ such that, for all $\alpha\in\N^3$, $0\leq \epsilon_1\leq D'$, 
\begin{equation}
\epsilon_1^{|\alpha|}\sum_{|\beta |\leq 2}\, \sup _{x\in\Omega '_{\epsilon_1}}\, \|\partial _x^\alpha a_\beta \|_{\cB_{2-|\beta |}}\ \leq \ C_p^{|\alpha |+1}\cdot (|\alpha |!)\period\label{eq:coeff-anal}
\end{equation}
We show that there exists $B>0$ such that, for all $\epsilon>0$, $j\in\N$, $r\in\{0;1;2\}$, and $\alpha\in\N^3$,
\begin{equation}
r+|\alpha |<2+j\ \impl \epsilon ^{r+|\alpha |}N_{j\epsilon , r}(D_x^\alpha \varphi )\ \leq \ B^{r+|\alpha |+1}\period\label{eq:borne-derivee-N}
\end{equation}
Take $B_0>0$ such that \eqref{eq:borne-derivee-N} holds true for $j\in\{0;1\}$ with $B=B_0$. 
We choose $B\geq \max (B_0, 2C_p\langle D'\rangle , C_a)$, where $C_a=1+\sharp \{(r,\beta)\in \{0; 1; 2\}\times \N^3; r+|\beta |<2\}$. Now we can follow the arguments in \cite{h1} (see also \cite{fhhs2}) to prove \eqref{eq:borne-derivee-N} by induction on $j$. As explained in \cite{h1}, $\varphi \in C^\omega (\Omega ';\cW _2)$ follows from \eqref{eq:borne-derivee-N} and \eqref{caract-anal} with $\delta =0$. 
 
\appendix
\section{Appendix}
\label{appendix}
\setcounter{equation}{0}

Here we explain the characterizations \eqref{caract-anal} and prove 
\eqref{eq:derivees-successives}. \\
In dimension $d=1$, the characterizations \eqref{caract-anal} are identical and well-known (cf.\ \cite{h3}). Let $d\geq 1$ and $u\in C^\infty(O;\cA)$. If $u$ is analytic then \eqref{caract-anal} holds true with $\delta =1$ (cf.\ \cite{h3}). This estimate implies \eqref{caract-anal} with $\delta =0$, since, by induction on $d$, there exists $M_d>0$ such that, for all $\alpha\in\N^d$, $(\alpha !)\leq M_d^{|\alpha |+1}(|\alpha |!)$. By \eqref{caract-anal} with $\delta =0$, $u$ is analytic in each variable, the others being kept fixed, yielding the analyticity of $u$ (cf.\ \cite{h3}).\\
Using Cauchy integral formula for analytic functions in several variables (cf.\ \cite{h3}), we prove here the following extension of \eqref{eq:derivees-successives}. For $d\in\N^\ast$, 
\begin{equation}\label{eq:derivees-successives-d}
\exists K>0\, ;\ \forall \alpha\in\N^d\comma\ \forall y\in\R^d\setminus\{0\}\comma\hspace{.4cm} \bigl|D^\alpha |\cdot |^{-1}\bigr|(y)\ \leq \ K^{|\alpha |+1}(\alpha !)\, |y|^{-|\alpha |-1}\period 
\end{equation}
In dimension $d=1$, one can show \eqref{eq:derivees-successives-d} with $K=1$ by induction. \\
Since $|\cdot |^{-1}$ is homogeneous of degree $-1$, $D^\alpha|\cdot |^{-1}$ is homogeneous of degree $-1-|\alpha|$, for all $\alpha$. Thus it suffices to prove \eqref{eq:derivees-successives-d} for $y$ in the unit sphere $\SSS ^d$ of $\R^d$. Let $\sqrt{\cdot}$ be the analytic branch of the square root that is defined on $\C\setminus\R ^{-}$. Take $y\in\SSS^d$. The well defined function $u:\cD\dans \{z\in\C\, ;\,  |z|\leq 4/\sqrt{7}\}$ given by 
\begin{eqnarray*}
 \cD \, = \, \bigl\{z=(z_1, \cdots , z_d)\in \C^d\, ; \forall 
j\comma\ |z_j|<(4\sqrt{d})^{-1}\bigr\}\comma\ u(z)\ =\ 
\frac{1}{\sqrt{\sum_{j=1}^d(y_j+z_j)^2}}\comma
\end{eqnarray*}
is analytic. By Cauchy inequalities (cf.\ Theorem 2.2.7, p. 27, in \cite{h3}), 
\begin{equation}\label{eq:borne-derivees-successives-d}
\forall \alpha\in\N^d\comma\ |\partial _z^\alpha u(0)|\ \leq\ 4\cdot 7^{-1/2}\cdot (\alpha !)\cdot ((4\sqrt{d})^{-1})^{-|\alpha|}\ \leq\ (4\sqrt{d})^{|\alpha |+1}(\alpha !)\period 
\end{equation}
Here $\partial _{z_j}:=(1/2)(\partial _{\Re z_j}+i\partial _{\Im z_j})$ but it can be replaced by $\partial _{\Re z_j}$ in the formula since $u$ is analytic. Now \eqref{eq:derivees-successives-d} follows from \eqref{eq:borne-derivees-successives-d} since, for all $\alpha$, 
\[(\partial _{\Re z}^\alpha u)(0)\ =\ i^{|\alpha|}(D^\alpha |\cdot |^{-1})(y)\period\]
%


%
%

\begin{thebibliography}{x}




%
\bibitem[CFKS]{cfks} H.L. Cycon, R.G. Froese, W. Kirsch, B. Simon: {\em 
Schr\"odinger operators with applications to quantum mechanics and 
global geometry.} Springer, 1987. 
%
\bibitem[FHHS1]{fhhs1} S. Fournais, M. Hoffmann-Ostenhof, T. Hoffmann-Ostenhof, T. \O stergaard S\o rensen: {\em The electron density is smooth away from the nuclei.} 
Comm. Math. Phys. {\bf 228}, no. {\bf 3} (2002), 401-415. 
%
\bibitem[FHHS2]{fhhs2} S. Fournais, M. Hoffmann-Ostenhof, T. Hoffmann-Ostenhof, T. \O stergaard S\o rensen: {\em Analyticity of the density of electronic wave functions.} 
Ark. Mat. {\bf 42}, no. {\bf 1} (2004), 87-106. 
%
\bibitem[FHHS3]{fhhs3} S. Fournais, M. Hoffmann-Ostenhof, T. Hoffmann-Ostenhof, T. \O stergaard S\o rensen: {\em Non-isotropic cusp conditions and regularity of the electron density of molecules at the nuclei.} Ann. Henri Poincaré {\bf 8} (2007), 731-748.
%
\bibitem[FH]{fh} R.G. Froese, I. Herbst: {\em Exponential bounds and 
absence of positive eigenvalues for $N$-body Schr\"odinger operators.} 
Comm. Math. Phys. {\bf 87}, no. {\bf 3}, 429-447 (1982).   
%
\bibitem[H\"o1]{h1}L. H\"ormander: {\em Linear partial differential operators.} Fourth printing Springer Verlag, 1976.
%
\bibitem[H\"o2]{h2}L. H\"ormander: {\em The analysis of linear partial differential operators 
III.} Springer Verlag, 1985.
%
\bibitem[H\"o3]{h3}L. H\"ormander: {\em An introduction to complex analysis in several variables.} Elsevier science publishers B.V., 1990.
%
\bibitem[Hu]{hu}W. Hunziker: {\em Distortion analyticity and molecular resonance curves.} Ann. Inst. H. Poincar\'e, section A, tome {\bf 45}, no {\bf 4}, p. 339-358 (1986). 
%
\bibitem[K]{k} T. Kato: {\em Pertubation theory for linear operators.} 
Springer-Verlag 1995.
%
\bibitem[KMSW]{kmsw} M. Klein, A. Martinez, R. Seiler, X.P. Wang: {\em On the Born-Oppenheimer expansion for polyatomic molecules.} Comm. Math. Phys.  {\bf 143}, no. {\bf 3}, 607-639 (1992). 
%
\bibitem[MS]{ms}A. Martinez, V. Sordoni: {\em Twisted pseudodifferential calculus and application to the quantum evolution of molecules.} Memoirs Am. Math. Soc., Vol. {\bf 200}, n. {\bf 936} (2009).
%
\bibitem[RS2]{rs2}M. Reed, B. Simon: {\em Methods of Modern Mathematical 
Physics, Vol. II : Fourier Analysis, Self-adjointness.} Academic Press, 1979.
%
\bibitem[RS4]{rs4}M. Reed, B. Simon: {\em Methods of Modern Mathematical 
Physics, Vol. IV : Analysis of operators.} Academic Press, 1979.
%

\end{thebibliography}
\end{document}